\documentclass[aps,floatfix,superscriptaddress,reprint,10pt,pra]{revtex4-1}
\usepackage{bm}
\usepackage{amsmath}
\usepackage{amssymb}
\usepackage{multirow}
\usepackage{empheq}
\usepackage{graphicx}
\usepackage{mathrsfs}
\usepackage{amsfonts}
\usepackage{amsthm}
\usepackage{color}
\usepackage{bigints}
\usepackage{txfonts}
\usepackage{hyperref}
\hypersetup{
	unicode=false,          
	pdftoolbar=true,        
	pdfmenubar=true,        
	pdffitwindow=false,     
	pdfstartview={FitH},    
	pdftitle={My title},    
	pdfauthor={Author},     
	pdfsubject={Subject},   
	pdfcreator={Creator},   
	pdfproducer={Producer}, 
	pdfkeywords={keyword1} {key2} {key3}, 
	pdfnewwindow=true,      
	colorlinks=false,       
	linkcolor=red,          
	citecolor=green,        
	filecolor=magenta,      
	urlcolor=cyan           
}
\makeatletter \tolerance = 10000 \tolerance = 10000
\usepackage{color}

\makeatother
\begin{document}

\title{Chirality-2 fermion induced Anti-Klein tunneling in 2D checkerboard lattice}

	\author{ Jiannan Hua }
     \affiliation{Department of Physics, School of Science, Westlake University, Hangzhou, Zhejiang 310024, China }
     \affiliation{Institute of Natural Sciences, Westlake Institute for Advanced Study, Hangzhou, Zhejiang 310024 , China}

	\author{Z. F. Wang}
	\affiliation{Hefei National Research Center for Physical Sciences at the Microscale, CAS Key Laboratory of Strongly-Coupled Quantum Matter Physics, Department of Physics, University of Science and Technology of China, Hefei, Anhui 230026, China }
	\affiliation{Hefei National Laboratory, University of Science and Technology of China, Hefei, Anhui 230088, China }

	\author{W. Zhu}
     \affiliation{Department of Physics, School of Science, Westlake University, Hangzhou, Zhejiang 310024, China }
     \affiliation{Institute of Natural Sciences, Westlake Institute for Advanced Study, Hangzhou, Zhejiang 310024 , China}
     
     \author{Weiwei Chen}
     \thanks{Corresponding author. E-mail: chenweiwei@cjlu.edu.cn}
     \affiliation{Key Laboratory of Intelligent Manufacturing Quality Big Data Tracing and Analysis of Zhejiang Province, College of Science, China Jiliang University, Hangzhou, 310018, China}

	\date{\today}

\begin{abstract}
The quantum tunneling effect in the two-dimensional (2D) checkerboard lattice is investigated. By analyzing the pseudospin texture of the states in a 2D checkerboard lattice based on the low-energy effective Hamiltonian, we find that this system has a chiral symmetry with chirality equal to 2. Although compared to regular chiral fermions, its pseudospin orientation does not vary uniformly. This suggests that the perfect reflection chiral tunneling, also known as the anti-Klein tunneling, is expected to appear in checkerboard lattice as well. In order to verify the conjecture, we calculate the transmission probability and find that normally incident electron states can be perfectly reflected by the barrier with hole states inside, and vice versa. Furthermore, we also numerically calculate the tunneling conductance of the checkerboard nanotube using the recursive Green's function method. The results show that a perfect on-off ratio can be achieved by confining the energy of the incident states within a certain range. It also suggests that, by tuning the barrier, the checkerboard nanotube is able to work as a perfect ``band filter" or ``tunneling field effect transistor", which transmits electrons selectively with respect to the pseudospin of the incident electrons.
\end{abstract}
	
\maketitle
	
\section{Introduction}

Quantum tunneling refers to the passage of particles with finite probability through barriers that are forbidden according to the laws of classical physics \cite{Huard2007,Gorbachev2008}. 
Nevertheless, quantum tunneling may bring serious problems when the size of transistors goes down to the nanoscale, e.g. logic errors occur if electrons start tunneling through the barriers when the transistor is off. 
Therefore, precisely controlling the quantum tunneling effect in the nanoscale transistor is of vital importance for the next-generation electronics \cite{Young2009NatPhy,Stander2009prl,Rutter2011NatPhy,Mak2009prl,Yuanbo2009Nature,Ohta2006Science,Zhang2008prb,Kuzmenko2009prb}.



Traditional transistors were designed relying on the fundamental charge degree of freedom of electrons, and then the intrinsic spin was also confirmed to modulate the electron transport, giving rise to the study of spintronics\cite{Zutic2004,Pulizzi2012,Awschalom2007}. Recently, other two degrees of freedom, valley and pseudospin, have been widely investigated in various quantum systems such as monolayer graphene, bilayer graphene and graphene-based heterojunctions, based on which several types of band filters have been proposed \cite{Wakabayashi2002,Bai,Nakabayashi,Schaibley2016,Vitale2018,Yu2020,McCann2009,Katsnelson2006}. The symmetry consideration is significant in these designing strategies, especially the chiral symmetry in Dirac fermions, which is deeply related to the perfect transmission and perfect reflection in nanostructures \cite{Wakabayashi2002,Nakabayashi,Habib2015prl,He2013,Killi2011prl}. 

In the case of systems with odd chirality, such as the monolayer graphene, the normally incident electrons can completely pass through a barrier of arbitrary height (known as the Klein paradox) \cite{Katsnelson2006,Bai2007prb,Gutierrez2016,Wang2008}. For systems with even chirality, such as the Bernal bilayer graphene \cite{Katsnelson2006,Gutierrez2016,Du2018prl}, the chiral nature leads to the opposite effect where electrons are always perfectly reflected for a sufficiently wide barrier for normal incidence, also known as the anti-Klein tunneling. However, the perfect reflection in the bilayer graphene is only achieved under two-band approximation since an interlayer bias breaks the pseudospin structure \cite{Duppen2013,Lu2015}, therefore the on-off ratio is low in these materials. Moreover, similar anti-Klein tunneling effects have also been reported in the spin-orbit systems and anisotropic electronic structures \cite{Anna2018,Ocampo2019}. Essentially, they all depend on the orientation of spin or pseudospin.


In this paper, we investigate the quantum tunneling in a 2D checkerboard lattice \cite{Montambaux,Wise2008NatPhy,Zeng2018npj,Wu2016,Sun2009prl,Sun2011prl}, where an anisotropic chiral symmetry exists. Analyzing the pseudospin texture based on the low-energy effective Hamiltonian, we find that the orientations of the pseudospin and wavevector in this system are locked, and when the angle of wavevector changes $2\pi$, the angle of pseudospin changes $2\times 2\pi$. Besides, the pseudospin textures of the Fermi surface above and below the touching point are centrosymmetric. These behaviors indicate the existence of chiral symmetry in this system, and the chirality is equal to 2, which suggest that the perfect reflection chiral tunneling is expected to appear in checkerboard lattice as well.

This conjecture of the perfect reflection in the checkerboard lattice is then confirmed by the calculation of transmission probability, which shows that a perfect reflection effect for normal incidence is found in the case of $s\neq s'$ where $s$ and $s'$ are the subband indexes inside and outside the barrier, respectively. Inspired by the perfect reflection of normal incidence in the checkerboard lattice, we also suppose that a ``band filter" or ``tunneling field effect transistor" can be designed based on the quasi-1 dimensional checkerboard nanotube. We numerical calculate the tunneling conductance of the checkerboard nanotube using the recursive Green's function method \cite{MacKinnon,Lewenkopf}. The results show that the current can be entirely blocked by the barrier potential in a certain range. Thus, by tuning the barrier, the checkerboard nanotube is able to work as a perfect ``band filter" or ``tunneling field effect transistor", which transmits electrons selectively with respect to the pseudospin of the incident electrons.

At last, we propose that $\tau$-type organic conductors \cite{Osada2019,Papavassiliou} and optical crystals \cite{Paananen2015} can serve as ideal platforms for creating functional digital devices made of checkerboard lattice and achieving perfect reflection.


\begin{figure}
	\centering
	\includegraphics[width=1.0\linewidth]{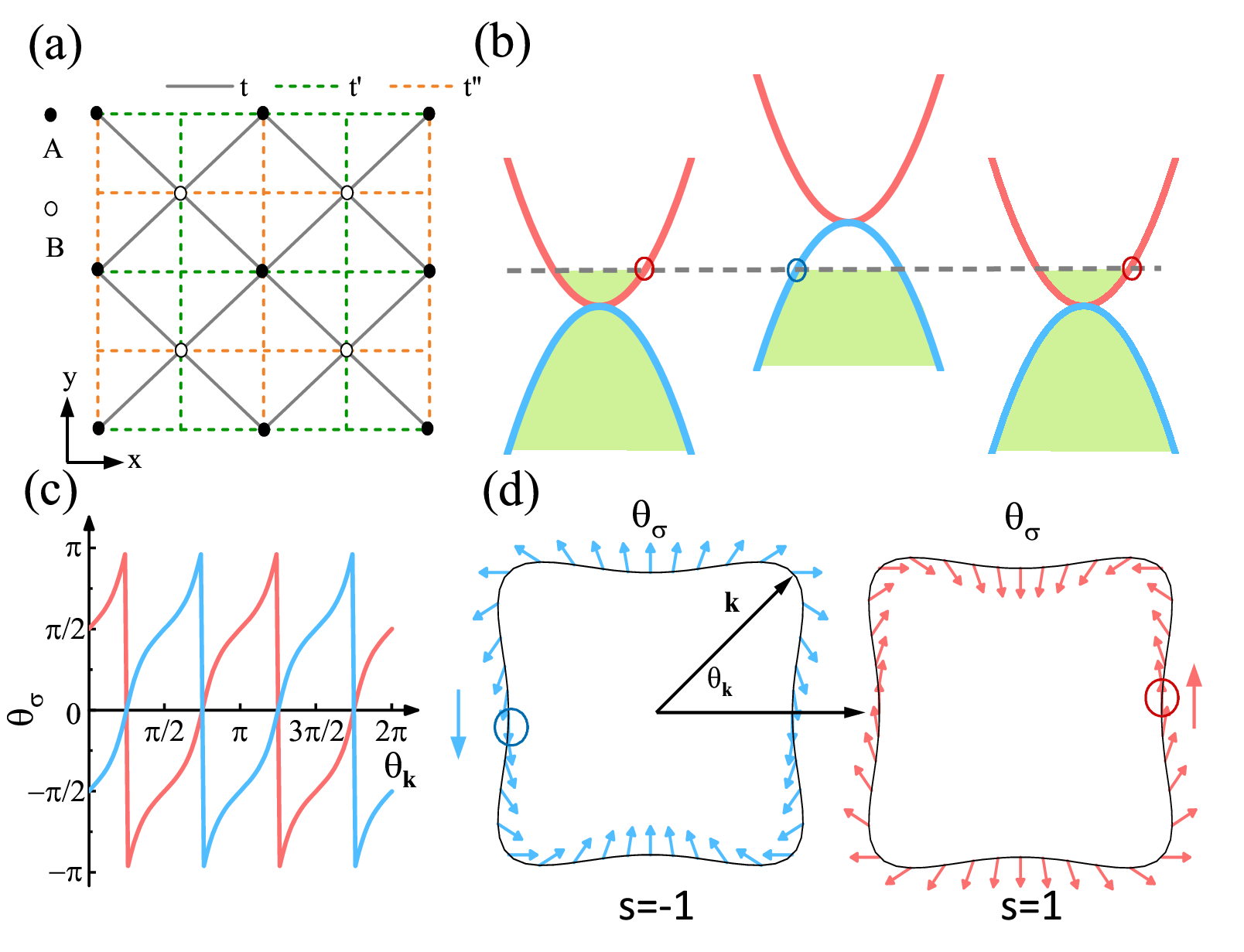}
	\caption{(a) Checkerboard lattice with nearest and next-nearest neighbor hoppings. Two sublattices are labeled by solid and open circles. (b) Schematic diagrams of the band structure and variation of the electrostatic energy caused by the barrier. (c) The angle of pseudospin as a function of the angle of wavevector: red for subband $s=1$ and blue for $s=-1$. (d) The textures of pseudospin at the Fermi surface with subband index $s=-1$ (blue) and $s=1$ (red) are denoted by the arrows of unit vectors.}
	\label{fig:chiralschematic}
\end{figure}

\section{Model}

We start with the tight-binding model of the checkerboard lattice depicted in Fig.~\ref{fig:chiralschematic}(a),
\begin{equation}\label{eq:ham}  
	\begin{aligned}
	H=&-\sum_{i,j}t(a^{\dagger}_{i,j}b_{i,j}+a^{\dagger}_{i,j}b_{i,j-1}+a^{\dagger}_{i,j}b_{i-1,j}+a^{\dagger}_{i,j}b_{i-1,j-1})\\
	&+t'(a^{\dagger}_{i,j}a_{i+1,j}+b^{\dagger}_{i,j}b_{i,j+1})
	+t''(a^{\dagger}_{i,j}a_{i,j+1}+b^{\dagger}_{i,j}b_{i+1,j})\\
	&+\text{H.c.}
	\end{aligned}
\end{equation}
where $a^{\dagger}_{i,j}(b^{\dagger}_{i,j})$ and $a_{i,j}(b_{i,j})$ are, respectively, the single electron creation and annihilation operators on the site $A(B)$ of the primitive cell $(i,j)$ with $i(j)$ being index along the $x(y)$-direction. 
$t$ stands for the nearest hopping, while $t'$ and $t''$ denote two types of next-nearest hopping. In the calculations below, without loss of generality, we choose the case with $t=t'=-t''=1$. A general analysis of parameter settings is provided in Appendix \ref{app:general model}.

Following the Bloch theorem, the Hamiltonian in the wavevector space reads (details shown in Appendix \ref{app:tight-binding})
\begin{equation}\label{eq:tilted Hamiltonian}
	\begin{aligned}
		H_{\tilde{\bm{k}}}&=
		(-2\cos \tilde{k}_x+2\cos \tilde{k}_y )\sigma_z-4\cos\frac{\tilde{k}_x}{2}\cos\frac{\tilde{k}_y}{2}\sigma_x
	\end{aligned}
\end{equation}
where $\bm{\sigma}=(\sigma_x,\sigma_y,\sigma_z)$ is the Pauli matrix of sublattice pseudospin.  The conduction and valence bands of this system quadratically touch at the $(\tilde{k}_x,\tilde{k}_y)=(\pi,\pi)$. Thus, we expand the above Hamiltonian around the touching point by redefining the wavevector as $\tilde{k}_x=\pi+k_x$ and $\tilde{k}_y=\pi+k_y$. The low-energy effective Hamiltonian is given by
\begin{equation}\label{eq:eff-H}
	H_{\bm{k}}=(k_x^2-k_y^2 )\sigma_z-k_xk_y\sigma_x
\end{equation}
The corresponding eigenenergy and eigen-state are
\begin{equation}
	E_{\bm{k}s}=sk^2\sqrt{\cos^22\theta_{\bm{k}}+\frac{1}{4}\sin^22\theta_{\bm{k}}}
\end{equation}
and 
\begin{equation}\label{eq:wavefunction primitive cell}
	|\psi_{\bm{k}s} \rangle =A_{\bm{k}s}\begin{pmatrix}
		\frac{1}{2}\sin2\theta_{\bm{k}} \\
           \cos2\theta_{\bm{k}}-\frac{E_{\bm{k}s}}{k^2}
	\end{pmatrix} 
\end{equation}
Here, 
$A_{\bm{k}s}=k^2 \left[2 E_{\bm{k}s}(E_{\bm{k}s}-k^2 \cos 2\theta_{\bm{k}} )\right]^{-1/2}$ 
is the normalization coefficient, $k=|\bm{k}|$ and $\theta_{\bm{k}}=\arctan(k_y/k_x )$ are the length and angle of $\bm{k}$, respectively, and $s=\pm1$ denotes different subbands. The pseudospin of this model occurs in the $(x-z)$ plane since the Hamiltonian Eq.~(\ref{eq:eff-H}) satisfies the anticommutation relation $\{H_{\bm{k}},\sigma_y\}=0$ \cite{footnote}. In order to describe the orientations of the pseudospin and the wavevector in the same plane, we perform a $\frac{\pi}{2}$-rotation along the $x$-direction in the pseudospin space and rewrite the Hamiltonian as
\begin{equation}
	H'_{\bm{k}}=e^{i\sigma_x\frac{\pi}{4}}H_{\bm{k}}e^{-i\sigma_x\frac{\pi}{4}}
	=k^2(\sigma_y\cos2\theta_{\bm{k}}-\frac{1}{2}\sigma_x\sin2\theta_{\bm{k}}).
\end{equation}
Therefore, the angle of pseudospin $\theta_{\bm{\sigma}}$ is obtained by solving
\begin{equation}\begin{aligned}
	\cos\theta_{\bm{\sigma}}=&\langle\psi'_{\bm{k}s}|\sigma_x|\psi'_{\bm{k}s}\rangle=\frac{s\sin2\theta_{\bm{k}}}{\sqrt{\sin^22\theta_{\bm{k}}+4\cos^22\theta_{\bm{k}}}}
	\\
	\sin\theta_{\bm{\sigma}}=&\langle\psi'_{\bm{k}s}|\sigma_y|\psi'_{\bm{k}s}\rangle=\frac{2s\cos2\theta_{\bm{k}}}{\sqrt{\sin^22\theta_{\bm{k}}+4\cos^22\theta_{\bm{k}}}}
\end{aligned}\end{equation}
where $|\psi'_{\bm{k}s} \rangle =e^{i\sigma_x\frac{\pi}{4}}|\psi_{\bm{k}s}\rangle$ is the rotated pseudospin state. It is remarkable that the pseudospin angle depends only on the angle of wavevector, not on the amplitude.

As shown in Fig.~\ref{fig:chiralschematic} (c) and (d), the angle of the pseudospin varies monotonically with the angle of the wavevector, and changes $2\times2\pi$ when the wavevector orientation changes $2\pi$. Besides, the pseudospin textures of the Fermi surface with energy $\pm E$ are centrosymmetric about the touching point. These behaviors are quite similar to chiral fermions with chirality equal to 2, such as the Bernal bilayer graphene. The difference is that in the bilayer graphene, the change of pseudospin angle is always twice the change of wavevector angle, while in the case of checkerboard, $\theta_{\bm{\sigma}}$ does not uniformly change with $\theta_{\bm{k}}$. This is related to the anisotropic Fermi surface of this model shown in Fig.~\ref{fig:chiralschematic} (d). 

In the normal incidence condition where $\theta_{\bm{k}}=0$ or $\pi$, this anisotropic pseudospin texture does not break the perfect reflection, also called the anti-Klein tunneling, reported in the regular 2-chiral fermion. As marked by the red and blue circles in Fig.~\ref{fig:chiralschematic} (b) and (d), when the states inside and outside the barrier belong to different subbands, the wavefunctions across the barrier are orthogonal due to the opposite pseudospin orientations, which leads to a perfect reflection if the barrier width tends to infinity, i.e. a potential step.


\begin{figure}
	\centering
	\includegraphics[width=\columnwidth]{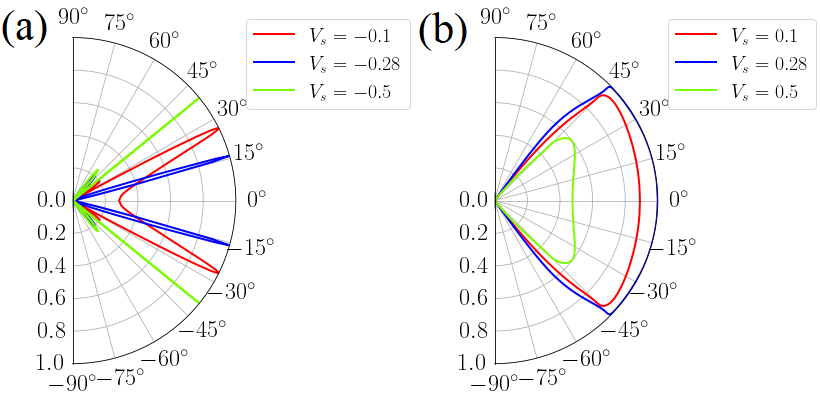}
	\caption{The angular dependence of the transmission probability in the checkerboard lattice. 
		We set the barrier potential (a) $V_s=-0.50, -0.28, -0.10$ and (b) $V_s=0.50, 0.28, 0.10$. 
		The incident energy is set to be $E=-0.1$.}
	\label{fig:transmission}	
\end{figure}

\section{Barrier potential and Transmission probability} 

Above, we infer the existence of the anti-Klein tunneling in the checkerboard lattice by analyzing the chiral symmetry and pseudospin texture of the low-energy effective Hamiltonian. In the following, we calculate the tunneling transmission probability to address this conjecture.
The barrier potential we considered is uniform along the $y$-direction and has a rectangular shape along the $x$-direction,
\begin{equation}
	V(x)=\left\{\begin{array}{ll}
		V_s,&0\le x\le D\\
		0,&\text{otherwise}
	\end{array}\right.
\end{equation}
where $V_s$ and $D$ denote the height and length of the barrier, respectively.
We assume that the incident wave comes from infinite away ($x\to-\infty$) with wavevector $\bm{k}=( k_x, k_y) $ and energy $E$ which satisfies the dispersion relation.
The wavefunctions of the system with barrier can be obtained by solving the equation (details shown in Appendix \ref{app:transmission})
\begin{equation}\label{eq:wave-equation-1}
	[\hat{H}_{\bm{k}}+V(x)]\psi(x,y)=E\psi(x,y),
\end{equation}
Here, $\hat{H}_{\bm{k}}$ serves as an operator in the coordinate representation. It is got by redefining the wavevector as $\bm{k}=\tilde{\bm{k}} - (\pi, \pi)$ in Eq.~(\ref{eq:tilted Hamiltonian}) followed with Fourier transformation between $k_x(k_y)$ and $x(y)$ \cite{note-fourier}.
Separating variables in Eq.~(\ref{eq:wave-equation-1}) results in $(E-V_{\alpha})^2=4(\cos k_{\alpha x}-\cos k_y)^2+16\sin^2(k_{\alpha x}/2)\sin^2(k_y/2)$, where the subscript $\alpha=L,M,R$ denotes the incident, barrier and transmitting regions, respectively. 
Mathematically, this equation gives two sets of roots, which are denoted by $\pm k_{\alpha x}$ and $\pm k_{\alpha x}^{\prime}$, and correspond to modes $e^{\pm ik_{\alpha x}x}$ and $e^{\pm ik_{\alpha x}^{\prime}x}$, respectively. Thus, without loss of generality, the solution to Eq.~(\ref{eq:wave-equation-1}) is in the form of
\begin{equation}\label{eq:wave-func}
\begin{aligned}
	\psi_{\alpha}(x,y)=e^{ik_yy} \times 
      &\left[
	\frac{a_\alpha e^{ik_{\alpha x}x} }{\sqrt{\left| v(k_{\alpha x}) \right| (1+|\zeta_{\alpha,1}|^2) }} \begin{pmatrix}1\\ \zeta_{\alpha,1}\end{pmatrix} \right. \\
	&+\frac{b_\alpha e^{-ik_{\alpha x}x} }{\sqrt{\left| v(-k_{\alpha x}) \right| (1+|\zeta_{\alpha,2}|^2) }} \begin{pmatrix}1\\ \zeta_{\alpha,2}\end{pmatrix} \\
     &+\frac{c_\alpha e^{ik^{\prime}_{\alpha x}x} }{\sqrt{\left| v(k^{\prime}_{\alpha x}) \right| (1+|\zeta_{\alpha,3}|^2) }} \begin{pmatrix}1\\ \zeta_{\alpha,3}\end{pmatrix} \\
     &\left. +\frac{d_\alpha e^{-ik^{\prime}_{\alpha x}x} }{\sqrt{\left| v(-k^{\prime}_{\alpha x}) \right| (1+|\zeta_{\alpha,4}|^2) }} \begin{pmatrix}1\\ \zeta_{\alpha,4}\end{pmatrix} 
	\right],
\end{aligned}
\end{equation}
where $v(k_{\alpha x})=\frac{1}{\hbar} \frac{\partial E}{\partial k_x}\big|_{k_x=k_{\alpha x}}$ is the $x$-component of quasiparticle velocity.
Then we attempt to determine the coefficients. 
Firstly, $b_R=0$ since it corresponds to an extra ``incident'' wave towards the barrier region.
Next, attention is turned to the wavenumbers in the incident and transmitting regions.
To be specific, it's easy to see that roots $k_{Lx}= k_{Rx}= k_x$ definitely are real, and therefore modes $e^{\pm ik_{L x}x}$ and $e^{\pm ik_{R x}x}$ are propagating. 
But mathematically, roots $k_{Lx}^{\prime}$ and $k_{Rx}^{\prime}$ could be either imaginary or real, which leads to some differences in physics. 
For the former case, i.e. $k_{Lx}^{\prime}$ and $k_{Rx}^{\prime}$ are imaginary just as in the bilayer graphene \cite{Katsnelson2006}, we denote positive real values $\kappa_{L x}=ik_{L x}^{\prime}$,  $\kappa_{Rx}=ik_{Rx}^{\prime}$ and modes $e^{\pm \kappa_{L x}x}$, $e^{\pm \kappa_{R x}x}$ are therefore evanescent.
The convergence of the wavefunction requires coefficients $d_L=c_R=0$.
In the latter case, i.e. $k_{Lx}^{\prime}$ and $k_{Rx}^{\prime}$ are real, modes $e^{\pm ik_{L x}^{\prime}x}$ and $e^{\pm ik_{R x}^{\prime}x}$ are propagating. 
Here we choose the signs of $k_{Lx}^{\prime}$ and $k_{Rx}^{\prime}$ to satisfy $v(-k_{Lx}^{\prime})>0$ and $v(k_{Rx}^{\prime})<0$, respectively. So that we can also set $d_L=c_R=0$ for the same reason as $b_R=0$.
So far, for both cases other coefficients can be obtained by the continuity conditions for both the wavefunctions and their derivatives.
The reflection ($R$) and transmission ($T$) probabilities satisfy $R+T=1$ according to the particle number conservation.
To calculate them, it is worth noting that the incident wave may be scattered into all propagating waves whose velocity components in the $x$-direction are away from the barrier region. 
Specifically speaking, if the math gives two propagating and two evanescent modes in the incident and transmitting regions, $R=|r|^2$ and $T=|t|^2$.
While if the math gives four propagating modes in the incident and transmitting regions, $R=|r|^2+|r'|^2$ and $T=|t|^2+|t'|^2$.
The settings abovementioned make sure that coefficients $r=b_L/a_L$, $r'=c_L/a_L$, $t=a_R/a_L$ and $t'=d_R/a_L$.


Fig.~\ref{fig:transmission} shows the transmission probability as a function of the angle of wavevector $\theta=\arctan(k_y/k_x)$ in the case of the incident hole-like state with energy $E=-0.1$. We see that the transmission probability is finite (even close to $T=1$ at some special $V_s$) and insensitive to the angle for barrier potential with hole-like states ($E-V_s<0$). On the contrary, for the barrier potential with electron-like states ($E-V_s>0$), the transmission probability is almost zero for the angle $\theta=0^{\circ}$, which corresponds to the angle of quasiparticle velocity $\phi=\arctan(\frac{\partial E/\partial k_y}{\partial E/\partial k_x})$ also being zero, i.e. normal incidence. It should be noticed that, under low energy conditions, these two angles, $\theta$ and $\phi$, are related by the identity $\tan\phi=\tan\theta\frac{2\tan^2\theta-1}{2-\tan^2\theta}$, so the range $\theta\in[-54.7^{\circ},54.7^{\circ}]$ corresponds to the range $\phi\in[-90^{\circ},90^{\circ}]$. As a consequence, $T$ is strictly 0 for $|\theta|\in(54.7^{\circ}, 90^{\circ}]$ since the velocity angle of incident wave exceeds $90^{\circ}$.

For the normal incidence case ($k_y=0$), we get the analytical form of the transmission probability,
\begin{equation}\label{eq:transmission-ky0}  
	T =\left\{\begin{aligned}
		&\frac{4k_x^2q_x^2}{4 k_x^2q_x^2 +(k_x^2-q_x^2)^2\sin^2(q_xD)},&\text{if}\  s=s'\\
		&\frac{4k_x^2q_x'^2}{4k_x^2q_x'^2 +(k_x^2+q_x'^2)^2\sinh^2(q_x'D)},&\text{if}\ 
		s\neq s'\end{aligned}\right.
\end{equation}
where $s=\text{sgn}(E)$, $s'=\text{sign}(E-V_s)$, $k_x = \arccos(1-\frac{|E|}{2})$, $q_x =\arccos(1-\frac{|E-V_s|}{2})$, and $q_x' =\text{arcosh}(1+\frac{|E-V_s|}{2})$.
When the $s$ index (electron-like for $s=+$ or hole-like for $s=-$) of the incident state is the same as the states contained in the barrier, Eq.~(\ref{eq:transmission-ky0}) implies that for certain $E$ and $V_s$, the transmission probability periodically oscillates with the barrier length $D$, driven by the ``$\sin^2(q_xD)$" term. From the physical perspective, constructive interference occurs when $q_xD=N\pi$ ($N\in \mathbb{Z}$), while destructive interference occurs when $q_xD=(N+\frac{1}{2})\pi$ ($N\in \mathbb{Z}$). However, when the $s$ index of the incident state is opposite to the states contained in the barrier, the transmission probability decays exponentially with the barrier length $D$. Thus, a perfect reflection effect for the normal incidence will be found in the case of $s\neq s'$. These results further confirm the existence of anti-Klein tunneling in this system. Furthermore, this perfect reflection behavior can be achieved in this model within a large window of barrier height than that in AB-stacked bilayer graphene. 
Once a four-band AB-stacked bilayer graphene is considered \cite{Snymann2007,McCann,Nilsson2006,McCann2013}, the perfect reflection can only be achieved with the barrier height where the two bands away from the Dirac point do not contribute, since an interlayer bias breaks the pseudospin structure \cite{Duppen2013,Lu2015}.

~\\
\begin{figure}
	\centering
	\includegraphics[width=0.9\columnwidth]{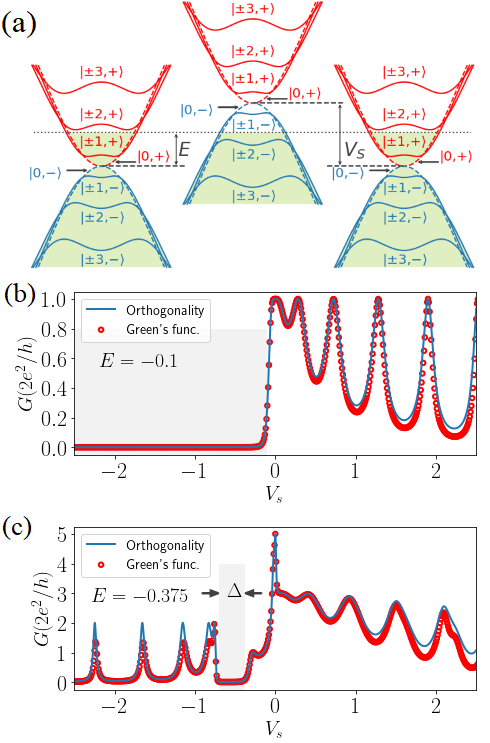} 
	\caption{
		The tunneling conductance of the checkerboard lattice varies with the barrier's height $V_s$. (a) The band structures of the tunneling barrier. The gray regions correspond to the situation that tunneling currents are almost entirely blocked.  Incident wave of different energy are calculated: (b) $E_1 =-0.1\in[-\Delta,0]$ and (c) $E_2 =-0.375<-\Delta$. Other parameters $M=10$, $\Delta=\sqrt{3}(1-\cos\frac{2\pi}{M})\approx0.33$ and $D=10$.
		The blue solid lines in (b) and (c) are results from the summation of independent transmission probabilities of different $k_y$ and the red dots are from recursive Green's function. }
	\label{fig:conductance}	
\end{figure}

\section{Tunneling Conductance} 

Inspired by the perfect reflection of the normal incidence in the checkerboard lattice, we suppose that a ``band filter" or ``tunneling field effect transistor" can be designed based on the quasi-1 dimensional checkerboard nanotube.
The nanotube is assumed to be infinity in the $x$-direction, but finite in the $y$-direction with a periodic boundary condition, so $k_y$ is discretized as $k_y=n\frac{2\pi}{M}$, where $M$ is the width of nanotube along $y$-direction and $n=\frac{M}{2},\frac{M}{2}-1,\cdots,0,\cdots,-\frac{M}{2}+1$. Here, we consider the lattice with even widths, which makes the band structure gapless. Then, we label each subband by $(n,s)$ as shown in Fig.~\ref{fig:conductance}~(a).
Two lowest bands with index $n=0$, which correspond to the normally incident states in the 2D checkerboard lattice, touch at $k_x=0$. Based on the band structure, we define the energy separation between the next-lowest band ($n=\pm1$) and the quadratic touching point as $\Delta$, and it can be calculated by the width of the lattice as $\Delta=\sqrt{3}(1-\cos\frac{2\pi}{M})$. 

Before calculating the conductance of the checkerboard nanotube with a barrier, we would like to analyze the orthogonality of the wavefunctions of different slices (assemblages of all primitive cells with the same $x$) in the nanotube since the wave propagates slice by slice. The slice wavefunction can be expressed by Eq.~(\ref{eq:wavefunction primitive cell}) as $\Psi_{n,s,k_x}(x)=\frac{1}{\sqrt{M}}[\psi_{n,s,k_x}(x,y_1),\cdots,\psi_{n,s,k_x}(x,y_M)]^\text{T}$, where $k_y$ is replaced by $n$. The wavefunctions of the two lowest bands take
\begin{equation}\begin{aligned}\label{eq:lowest-wavefunction}  
		\psi_{0,+,k_x}(\bm{r})=& 
		\begin{pmatrix}
			1 \\  0  \\
		\end{pmatrix} e^{i k_x x},\\
		\psi_{0,-,k_x'}(\bm{r})=& 
		\begin{pmatrix}
			0 \\  1  \\
		\end{pmatrix} e^{i k_x' x}.
\end{aligned}\end{equation}
where $\psi_{0,+,k_x}(\bm{r})$ only populates the sublattice A and $\psi_{0,-,k_x'}(\bm{r})$ only populates the sublattice B, which is similar to the zero-mode solution of the Dirac fermions in a magnetic field \cite{Neto2009}. Thus, one can get
\begin{equation}\label{eq:orthogonal-1}
	\Psi^{\dagger}_{0,+,k_x}(x)\Psi_{0,-,k_x'}(x)= \sum_{y}\psi^{\dagger}_{0,+,k_x}(\bm{r})\psi_{0,-,k_x'}(\bm{r})=0.
\end{equation}
Besides, the slice wavefunctions with different band index $n$ are also orthogonal due to
\begin{equation}\label{eq:orthogonal-2}
	\Psi^{\dagger}_{n,s,k_x}(x)\Psi_{n',s',k_x'}(x)\propto\sum_{y}e^{i(k_y-k_y')y}=\delta_{k_y,k_y'}=\delta_{n,n'}.
\end{equation}
Combining the relations Eq.~(\ref{eq:orthogonal-1}) and Eq.~(\ref{eq:orthogonal-2}), it is easy to see that slice wavefunction $\Psi_{0,s,k_x}(x)$ is orthogonal to all other slice wavefunctions $\Psi_{n,s',k_x'}(x)$ with $s'\ne s$. In other words, slice wavefunctions with energies in the range $E\in[0,\Delta]$ ($E\in[-\Delta,0]$) are orthogonal to those in the range $E<0$ ($E>0$), which implies that an incident electron-like wave with energy $E\in[0,\Delta]$ can not tunnel through a potential barrier with hole-like states inside, and similarly, an incident hole wave with energy $E\in[-\Delta,0]$ can not tunnel through a potential barrier with electron states inside. This is consistent with the anti-Klein tunneling we find under the normal incidence condition. Besides, it is worth to point out that for the evanescent wave, i.e. $k_x$ is imaginary, although the state does not appear in the dispersion relation Fig.~\ref{fig:conductance}~(a), orthogonality relations Eq.~(\ref{eq:orthogonal-1}) and Eq.~(\ref{eq:orthogonal-2}) still hold, as long as replace the subscript $n$ with corresponding transverse wavenumber $k_y=n\frac{2\pi}{M}$.

In order to verify the perfect reflection in the checkerboard nanotube, we calculate the tunneling conductance in two ways. One is to sum transmission probabilities over all channels obtained in the previous section, i.e. $G\approx\frac{2e^2}{h}\sum_{n=-M/2}^{M/2-1}T_n$, where $n$ represents the corresponding transverse wavenumber $k_y$ as previously mentioned. The other is to use the zero-bias Landauer formula combined with recursive Green's function method \cite{MacKinnon,Lewenkopf}. 

Fig.~\ref{fig:conductance} also reports results of the tunneling conductance $G$ versus the height of barrier $V_s$ with transverse width $M=10$, length $D=10$, and incident energies: (b) $E_1=-0.1$ and (c) $E_2=-0.375$. In this condition, the energy separation between the band $n=\pm1$ and $n=0$ is $\Delta\approx0.33$, so the incident state with energy $E_1\in[-\Delta,0]$ contains only contribution from band $(0,-)$ while the incident state with energy $E_2<-\Delta$ also contains contribution from band $n\ne0$. 

For the incident hole-like state with energy $E_1$, as shown in Fig.~\ref{fig:conductance}(b), the current is almost entirely blocked by the barrier potential when $E_1-V_s>0$, in which the barrier contains electron-like states inside, and perform resonant tunneling in other range.  For the incident energy $E_2$, shown in Fig.~\ref{fig:conductance}(c), the current is blocked by the barrier with height $V_s \in ( E_2-\Delta, E_2)$, which contains states from band $(0,+)$, and tunnels through the barrier resonantly in other range. These results are in good agreement with our predictions from orthogonality analysis of wavefunctions, and imply that a barrier potential in the checkerboard lattice can play the role of a ``band filter": when chemical potential $E$ is tuned to be in the range $[-\Delta,0]$, the negative barrier potential blocks the electron-like states tunneling while the positive barrier transmits these states. In addition, the peaks and valleys in Fig.~\ref{fig:conductance}(b) reflect the transmission enhancement from the resonances due to the constructive interference and the transmission suppression from the anti-resonances due to the destructive interference inside the barrier, respectively. This is consistent with the ``$\sin^2(q_xD)$" term in the case $s=s'$ of Eq.~(\ref{eq:transmission-ky0}).

From the comparison of results from these two methods, it can be seen that the tunneling conductance is insensitive to the bands coupling, especially in the perfect reflection region, which meets the expectation from the slice wavefunction analysis that $\Psi_{0,\pm,k_x}(x)$ and $\Psi_{n,\mp,k_x'}(x)$ are completely orthogonal. Another remarkable behavior found in tunneling conductance is the presence of many resonance peaks during the change of $V_s$, at which the barrier is transparent to one or more channels. These resonance peaks arise from waves that are reflected multiple times in the barrier and then transmitted in the same phase, which is similar to the taking place in the optical Fabry-Perot resonator or in a microwave capacitively-coupled transmission-line resonator \cite{Mahan2009}. This can be proven by that, in Fig.~\ref{fig:conductance}(b), the locations of resonance peaks are well matched to the resonance condition of transmission probability for the normal incidence, i.e., $q_xD=\pi N$.

\section{Materials realization} 

So far we have explored the perfect reflection Klein tunneling in the checkerboard lattice based on the tight-binding model. Then, we suggest some experimental systems where our simulation results can be potentially observed. At first, attention can be paid to the $\tau$-type organic conductors, in each conducting layer of which, donor molecules form a square lattice, and anion molecules are arranged on it with a checkerboard pattern \cite{Osada2019,Papavassiliou}. The fact that the conduction and valence bands exhibit the quadratic band touching at the corner of the square Brillouin zone was also confirmed by the tight-binding model and DFT calculations. Besides, the optical checkerboard-like lattices with cold atoms are also compelling candidates to simulate this condensed-matter problem due to the simple tuning of the parameters \cite{Paananen2015}. Lattice constants of these material candidates $a\lesssim 1nm$. From the results shown in Fig.~\ref{fig:conductance} and Fig.~\ref{fig:conductance-appendix} (in Appendix \ref{app: tunneling conductance}), it is evident that when the gate width is less than $10$ times lattice constants, the transmission in the anti-Klein region already approaches zero to an extreme extent, indicating that the contribution from evanescent waves is almost negligible. Therefore, it is obvious that using gate widths comparable to or even smaller than existing tunneling field effect transistors ($\gtrsim 10nm$) \cite{Lee2015,Hwang2019} can completely achieve perfect reflection.

\section{Conclusion and discussion}

In summary, we have studied the electronic quantum tunneling of a checkerboard lattice through a barrier potential.
Due to the chiral nature of the quasiparticles, we find that there exists an  anti-Klein tunneling, which leads to the perfect reflection of the normally incident waves. 
Moreover, we have also shown that a barrier potential can play the role of a
``band filter" or ``tunneling field effect transistor" in the checkerboard nanotube, which transmits the electronic states according to the selection rule.  
Finally, we expect that the checkerboard lattice can be realized in materials like $\tau$-type organic conductors and optical checkerboard-like lattices.

\textit{Acknowledgments.---}	
This work was supported by ``Pioneer" and ''Leading Goose" R\&D Program of Zhejiang (2022SDXHDX0005), the Key R\&D Program of Zhejiang Province (2021C01002).


\begin{widetext}

\begin{appendix}

\section{Tight-binding model}\label{app:tight-binding}

The tight-binding model of the checkerboard lattice depicted in Fig.~\ref{fig:chiralschematic}(a) is
\begin{equation}\label{eq:hamA}  
	\begin{aligned}
	H=&-\sum_{i,j}t(a^{\dagger}_{i,j}b_{i,j}+a^{\dagger}_{i,j}b_{i,j-1}+a^{\dagger}_{i,j}b_{i-1,j}+a^{\dagger}_{i,j}b_{i-1,j-1})\\
	&+t'(a^{\dagger}_{i,j}a_{i+1,j}+b^{\dagger}_{i,j}b_{i,j+1})
	+t''(a^{\dagger}_{i,j}a_{i,j+1}+b^{\dagger}_{i,j}b_{i+1,j})\\
	&+\text{H.c.}
	\end{aligned}
\end{equation}
where $a^{\dagger}_{i,j}(b^{\dagger}_{i,j})$ and $a_{i,j}(b_{i,j})$ are, respectively, the single electron creation and annihilation operator on the site $A(B)$ of the primitive cell $(i,j)$ with $i(j)$ being index along the $x(y)$-direction. 
$t$ stands for the nearest hopping, while $t'$ and $t''$ denote two types of next-nearest hopping. To get the Hamiltonian in the wavevector space, apply the Fourier transform 
    \begin{equation}\label{}  
    \begin{aligned}
        \hat{c}_{i,j} &= \frac{1}{\sqrt{\mathcal{V}}}\sum_{\tilde{\bm{k}}} e^{i\tilde{\bm{k}}\cdot \bm{r}_{i,j}^c} \hat{c}_{\tilde{\bm{k}}}      \\
    \end{aligned}
    \end{equation}
where $\mathcal{V}$ is the area of the lattice, and $c$ represents either $a$ or $b$, $\bm{r}_{i,j}^c$ is the position of the corresponding sublattice of the primitive cell $(i,j)$. In this letter, we take the length of primitive translation vectors as the unit length, which is also the distance of the next-nearest hopping. Then $\bm{r}_{i,j}^a=(i,j)$, $\bm{r}_{i,j}^b=(i+\frac{1}{2},j+\frac{1}{2})$. Each item in Eq.~(\ref{eq:hamA}) is in the form
\begin{equation}\label{}  
	\begin{aligned}
	\sum_{i,j} \hat{c}^{\dagger}_{i,j} \hat{c}^{\prime}_{i+\delta i,j+\delta j} &= \frac{1}{\mathcal{V}}\sum_{i,j}  \sum_{\tilde{\bm{k}},\tilde{\bm{k}}^\prime} e^{-i\tilde{\bm{k}}\cdot \bm{r}_{i,j}^c} \hat{c}_{\tilde{\bm{k}}}^\dagger e^{i\tilde{\bm{k}}^\prime \cdot \bm{r}_{i+\delta i,j+\delta j}^{c^\prime}} \hat{c}^{\prime}_{\tilde{\bm{k}}^\prime}  \\
             &= \frac{1}{\mathcal{V}} \sum_{\tilde{\bm{k}},\tilde{\bm{k}}^\prime} \sum_{i,j}  e^{i(\tilde{\bm{k}}^\prime- \tilde{\bm{k}} )\cdot \bm{r}_{i,j}^c }   e^{i\tilde{\bm{k}}\prime\cdot \delta \bm{r}} \hat{c}_{\tilde{\bm{k}}}^\dagger\hat{c}^{\prime}_{\tilde{\bm{k}}^\prime} \\
             &= \sum_{\tilde{\bm{k}}}   e^{i\tilde{\bm{k}}\cdot \delta \bm{r}} \hat{c}_{\tilde{\bm{k}}}^\dagger \hat{c}^{\prime}_{\tilde{\bm{k}}} \\
	\end{aligned}
\end{equation}
where  $\delta \bm{r}=\bm{r}_{i+\delta i,j+\delta j}^{c^\prime} - \bm{r}_{i,j}^c$. Thus, Eq.~(\ref{eq:hamA}) becomes

\begin{equation}\label{}  
	\begin{aligned}
	\hat{H}
      =&- \sum_{\tilde{\bm{k}} }\Big\{ t \left[ e^{i\tilde{\bm{k}} \cdot (\frac{1}{2}, \frac{1}{2})} + e^{i\tilde{\bm{k}} \cdot (\frac{1}{2}, -\frac{1}{2})}  + e^{i\tilde{\bm{k}} \cdot (-\frac{1}{2}, \frac{1}{2}) } + e^{i\tilde{\bm{k}} \cdot (-\frac{1}{2}, -\frac{1}{2}) }   \right]\hat{a}_{\tilde{\bm{k}} }^\dagger \hat{b}_{\tilde{\bm{k}} }  \\
       & + \left[ t' e^{i\tilde{\bm{k}} \cdot (1,0) } + t''e^{i\tilde{\bm{k}} \cdot (0,1) }  \right] \hat{a}_{\tilde{\bm{k}} }^\dagger \hat{a}_{\tilde{\bm{k}} } +  \left[ t'' e^{i\tilde{\bm{k}} \cdot (1,0) } + t'e^{i\tilde{\bm{k}} \cdot (0,1) }  \right] \hat{b}_{\tilde{\bm{k}} }^\dagger \hat{b}_{\tilde{\bm{k}} } \Big\} \\
       &+\text{H.c.}  \\
      =&- \sum_{\tilde{\bm{k}} } 4 t \cos \frac{\tilde{k}_x}{2} \cos\frac{\tilde{k}_y}{2} ( \hat{a}_{\tilde{\bm{k}} }^\dagger \hat{b}_{\tilde{\bm{k}} }+\text{H.c.} )  \\
       & + 2 ( t' \cos \tilde{k}_x  + t''\cos \tilde{k}_y ) \hat{a}_{\tilde{\bm{k}} }^\dagger \hat{a}_{\tilde{\bm{k}} } +2( t''\cos \tilde{k}_x + t'\cos \tilde{k}_y) \hat{b}_{\tilde{\bm{k}} }^\dagger \hat{b}_{\tilde{\bm{k}} }  \\
      =& \sum_{\tilde{\bm{k}} } \hat{\bm{c}}_{\tilde{\bm{k}} }^{\dagger} H_{\tilde{\bm{k}} } \hat{\bm{c}}_{\tilde{\bm{k}}} \\
	\end{aligned}
\end{equation}
where $\hat{\bm{c}}_{\tilde{\bm{k}} } = \begin{pmatrix} \hat{a}_{\tilde{\bm{k}} } \\ \hat{b}_{\tilde{\bm{k}}} \end{pmatrix}$, $H_{\tilde{\bm{k}}} = -2\begin{pmatrix}
		t'\cos \tilde{k}_x+t''\cos \tilde{k}_y&2t\cos\frac{\tilde{k}_x}{2}\cos\frac{\tilde{k}_y}{2}\\
		2t\cos\frac{\tilde{k}_x}{2}\cos\frac{\tilde{k}_y}{2}&t''\cos \tilde{k}_x+t'\cos \tilde{k}_y
	\end{pmatrix} $.

\section{Parameters analysis and $\tau$-type organic conductor}\label{app:general model}
As shown in Appendix \ref{app:tight-binding}, The wavevector space form of the Hamiltonian Eq.~(\ref{eq:ham}) with arbitrary parameters $t$, $t'$ and $t''$ is given by
\begin{equation}\begin{aligned}\label{app:H}
	H_{\tilde{\bm{k}}}=&-2\begin{pmatrix}
		t'\cos\tilde{k}_x+t''\cos\tilde{k}_y&2t\cos\frac{\tilde{k}_x}{2}\cos\frac{\tilde{k}_y}{2}\\
		2t\cos\frac{\tilde{k}_x}{2}\cos\frac{\tilde{k}_y}{2}&t''\cos\tilde{k}_x+t'\cos\tilde{k}_y
	\end{pmatrix}\\
	=&-(t'+t'')(\cos\tilde{k}_x+\cos\tilde{k}_y)\sigma_0-(t'-t'')(\cos\tilde{k}_x-\cos\tilde{k}_y)\sigma_z\\
	&-4t\cos\frac{\tilde{k}_x}{2}\cos\frac{\tilde{k}_y}{2}\sigma_x
\end{aligned}\end{equation}
where $\sigma_0$ is the identity matrix.
The corresponding dispersion relation is
\begin{equation}\begin{aligned}\label{app:Ek}
	E_{\tilde{\bm{k}}\pm}=&-(t'+t'')(\cos\tilde{k}_x+\cos\tilde{k}_y)\pm\\
	&\sqrt{(t'-t'')^2(\cos\tilde{k}_x-\cos\tilde{k}_y)^2+16t^2\cos^2\frac{\tilde{k}_x}{2}\cos^2\frac{\tilde{k}_y}{2}}
\end{aligned}\end{equation}

The parameters $t=t'=-t''=1$ set in the main text is a showcase without loss of generality, based on following reasons. Firstly, since the $\sigma_0$ term in the Hamiltonian Eq.~(\ref{app:H}) does not affect the expression of the eigenstates, the $t'=-t''$ we set do not affect the orientation of the pseudospin, which is essential reason of the Klein and anti-Klein tunneling. Secondly, from the dispersion relation Eq.~(\ref{app:Ek}), it is obvious that the location of the touching point $\tilde{\bm{k}}=(\pi,\pi)$ is independent with the values of $t$, $t'$ and $t''$. Thirdly, we performed the calculation of the pseudospin texture of the model with parameters are obtained by DFT calculation of $\tau$-type organic conductor reported in Ref.~\cite{Osada2019}, which gives $t=0.16$eV, $t'=0.13t$ and $t''=-0.07t$. As shown in Fig.~\ref{fig:organicmodel} (b) and (c), we find that when the orientation of wavevector changes $2\pi$ in the ($x$-$y$) plane, the orientation of pseudospin changes $2\times 2\pi$ in the ($x$-$z$) plane. In Fig.~\ref{fig:organicmodel} (b) and (c), we coincide the $z$-axis of pseudospin with the $y$-axis of wavevector. This implies that the values of $t$, $t'$ and $t''$ can affect the energy level of the touching point and the shape of the Fermi surface, but do not change the nature of the system, which is a chirality-2 fermion. Finally, for the normal incidence $\tilde{k}_y=\pi$, the off-diagonal components vanish. Hence, the system hold two $\tilde{k}_x$-independent orthogonal eigenstates $\begin{pmatrix}0\\1\end{pmatrix}$ and $\begin{pmatrix}1\\0\end{pmatrix}$ same as Eq.~(\ref{eq:lowest-wavefunction}), which directly induce the anti-Klein tunneling.

\begin{figure}
	\centering
     \includegraphics[width=0.5\linewidth]{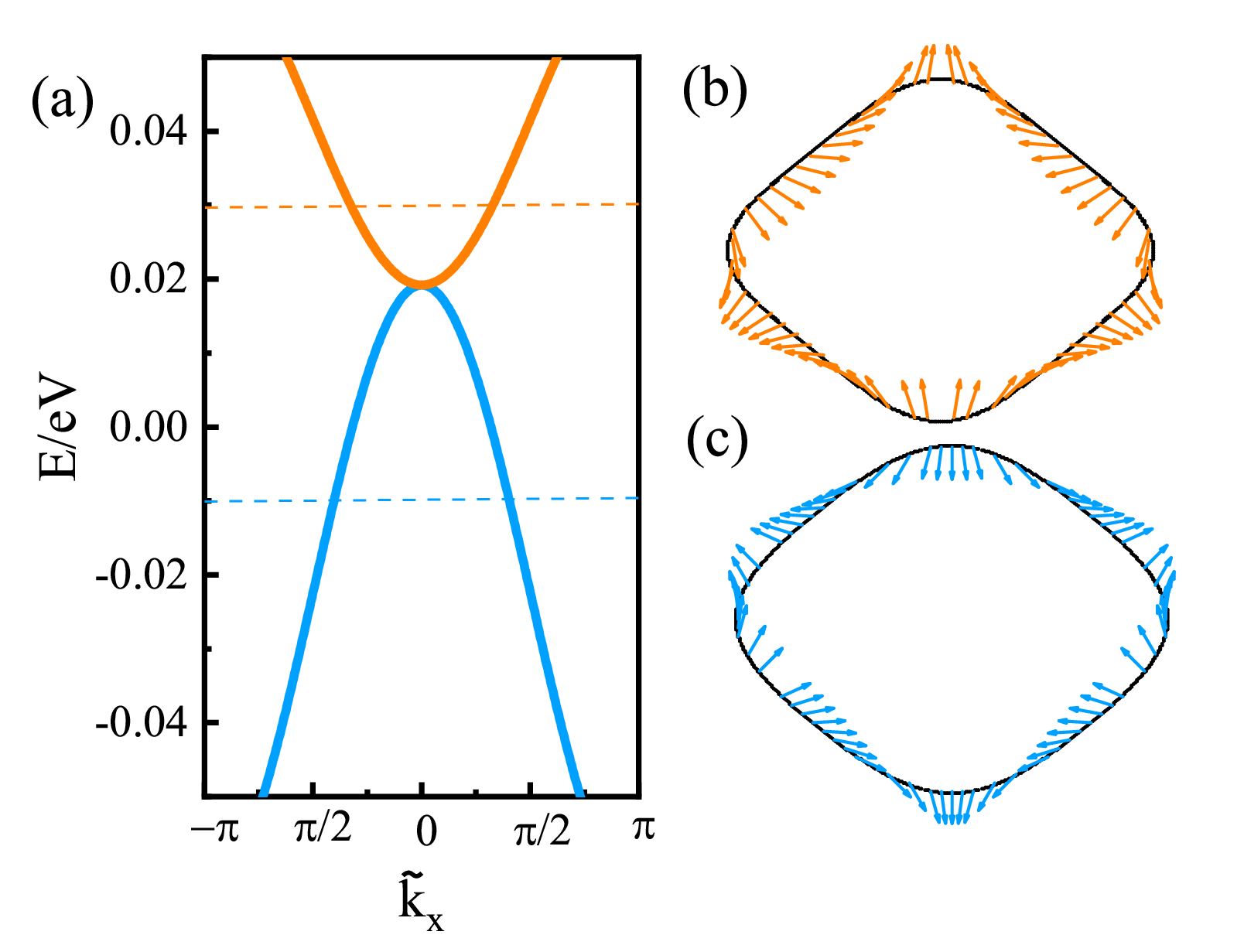}
	\caption{(a) Band structure of the Hamiltonian with parameters obtained by DFT fitting of $\tau$-type organic conductor. We set $\tilde{k}_y=\pi$. (b) and (c) Textures of pseudospin at the Fermi surfaces $E_f=0.03$eV (orange) and $E_f=-0.01$ eV (blue), respectively.}
	\label{fig:organicmodel}
\end{figure}

\section{Calculation of transmission probability}\label{app:transmission}

\subsection{Determining the wave modes}\label{Cases of tilted lattice}


Before solving Eq.~(\ref{eq:wave-equation-1}) globally, we need to determine the modes of wavefunction in each region.
Without loss of generality, we suppose the wavefunction as $\psi(x,y)=\begin{pmatrix}
           \zeta_A    \\
           \zeta_B   \\
           \end{pmatrix} e^{\lambda_x x} e^{\lambda_y y}$
and substitute it into Eq.~(\ref{eq:wave-equation-1})
    \begin{equation}\label{eigen-func-tiltde}  
           \hat{H}_{\bm{k}} \begin{pmatrix}
           \zeta_A    \\
           \zeta_B   \\
           \end{pmatrix} e^{\lambda_x x} e^{\lambda_y y} =  E^\prime \begin{pmatrix}
           \zeta_A    \\
           \zeta_B   \\
           \end{pmatrix} e^{\lambda_x x} e^{\lambda_y y},
    \end{equation}
where $E^\prime=E-V$ and $\hat{H}_{\bm{k}}$ 
is got by redefining the wavevector as $\bm{k}=\tilde{\bm{k}} - (\pi, \pi) $ in Eq.~(\ref{eq:tilted Hamiltonian}) followed with Fourier transformation between $k_x(k_y)$ and $x(y)$ \cite{note-fourier} to serve as an operator in the coordinate representation. 
Therefore, we get
    \begin{equation}\label{eigen_eq_tilted}  
    \begin{aligned}
        \begin{pmatrix}
            2\left[ \cos ( i \lambda_x ) -\cos ( i  \lambda_y )\right]   &   -4 \sin \frac{i  \lambda_x}{2}  \sin \frac{i  \lambda_y}{2}  \\
           -4   \sin \frac{i  \lambda_x}{2}  \sin  \frac{i  \lambda_y}{2}  &   2\left[ -\cos ( i  \lambda_x  ) + \cos ( i  \lambda_y  ) \right]   \\
        \end{pmatrix}
           \begin{pmatrix}
           \zeta_A     \\
           \zeta_B   \\
           \end{pmatrix} e^{\lambda_x x} e^{\lambda_y y}   &= E^\prime 
           \begin{pmatrix}
           \zeta_A     \\
           \zeta_B   \\
           \end{pmatrix} e^{\lambda_x x} e^{\lambda_y y}  \\
    \end{aligned}
    \end{equation}
Calculating 
    \begin{equation}\label{}  
    \begin{aligned}
        \det \begin{vmatrix}
         2\left[ \cos ( i \lambda_x ) -\cos ( i  \lambda_y )\right]- E^\prime    &   -4 \sin \frac{i  \lambda_x}{2}  \sin \frac{i  \lambda_y}{2}  \\
           -4   \sin \frac{i  \lambda_x}{2}  \sin  \frac{i  \lambda_y}{2}  &   2\left[ -\cos ( i  \lambda_x  ) + \cos ( i  \lambda_y  ) \right]   - E^\prime  \\
           \end{vmatrix}  = 0
    \end{aligned}
    \end{equation}
leads to
    \begin{equation}\label{Eprime2(lambda_x, lambda_y) (tilted)}  
        E^{\prime 2} = 4 \left( u - v \right)^2   +  4 \left( 1 - u \right) \left( 1 - v \right)  \\
    \end{equation}
where auxiliary values
    \begin{equation}\label{}  
    \begin{aligned}
        u \equiv \cos ( i \lambda_x ) &= 
            \begin{cases}
            \cos  k_x  \text{ for } \lambda_x = ik_x  \text{ is imaginary }\\
            \cosh k_x   \text{ for } \lambda_x=k_x \text{ is real }
            \end{cases} \\
        v \equiv \cos ( i  \lambda_y ) &= \cos  k_y  \text{ for } \lambda_y = ik_y \text{ is imaginary}\\
    \end{aligned}
    \end{equation}
Rewrite Eq.~(\ref{Eprime2(lambda_x, lambda_y) (tilted)}) as a quadratic equation with $u$ as the variable:
    \begin{equation}\label{}  
    u^2 - \left( 1+v \right) u + \left(v^2 - v + 1 - \frac{E^{\prime 2}}{4} \right) = 0
    \end{equation}
The solution is
    \begin{equation}\label{}  
    u_{\pm} = \frac{1+v \pm \sqrt{E^{\prime 2} - 3\left(1-v \right)^2}}{2}
    \end{equation}
Besides, Eq.~(\ref{Eprime2(lambda_x, lambda_y) (tilted)}) leads to that the necessary and sufficient condition for $\frac{\partial E^{\prime 2}}{\partial u} > 0$ is $u> \frac{1+v}{2}$.
As a consequence, mathematically, for a certain $v$, $E^{\prime 2}$ is monotonically decreasing for $u \in \left(-\infty, \frac{v+1}{2} \right)$, and monotonically increasing for $u \in \left(\frac{v+1}{2}, +\infty \right)$, respectively. Thus,  $E^{\prime 2}$ achieve minimum value $3(1-v)^2$ when $u = \frac{v+1}{2}$. Besides, $E^{\prime 2}=4(1-v)^2$ for $u=1$. These results in, physically, for $\left| E^{\prime}\right| < \sqrt{12}$ and a fixed $v \in [-1,1]$:

    \begin{itemize}
    \item If $ \left| E^{\prime}\right| < \sqrt{3}\left(1-v \right) $, there is not any mode.
    \item If $ \sqrt{3}(1-v ) < | E^{\prime}| <  2(1-v )$, just shown as the blue line in the Fig.~\ref{energy_band_tilted_E2.5}, there are four modes with
    \begin{equation}\label{}  
        \begin{cases}
            \lambda_x =  \pm ik_x = \pm i \arccos \left(u_{-}\right), \text{\quad propagating}\\
            \lambda_x =  \pm ik_x = \pm i \arccos \left(u_{+}\right), \text{\quad propagating}
        \end{cases} \\
    \end{equation}
And the transmission probability is easy to be calculated as introduced in Appendix \ref{General situation of two propagating modes}.
    \item If $ \left| E^{\prime}\right| >  2\left(1-v \right)$, just shown as the orange line in the Fig.~\ref{energy_band_tilted_E2.5}, there are four modes with
    \begin{equation}\label{}  
        \begin{cases}
            \lambda_x =  \pm ik_x = \pm i \arccos \left(u_{-}\right), \text{\quad propagating}\\
            \lambda_x =  \pm k_x = \pm \operatorname{arcosh} \left(u_{+}\right), \text{\quad evanescent}
        \end{cases} \\
    \end{equation}
And the transmission probability is easy to be calculated as introduced in Appendix \ref{General situation of one plane wave and one evanescent wave}.
    \end{itemize}

	\begin{figure}[!htbp]
	\centering
	\includegraphics[scale=0.5]{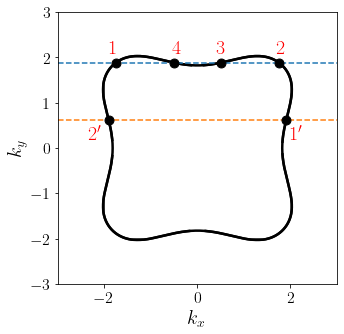}
	\caption{The black curve is a typical Fermi surface of the checkerboard lattice for $t=t^\prime=-t''=1$. Specifically here $E^\prime=2.5$. The orange and blue lines correspond to $k_y=\frac{3\pi}{5}$ and $\frac{\pi}{5}$, respectively. The black curve and the orange line intersect at points $1^\prime$ and $2^\prime$ which represent propagating modes of the same group. The black curve and the blue line intersect at points 1, 2, 3 and 4, which represent propagating modes belong to two groups.}
	\label{energy_band_tilted_E2.5}
	\end{figure}

So far, for certain $E^\prime = E-V$ and $k_y$, Eq.~(\ref{eq:wave-equation-1}) gives two sets of roots in each region, which are denoted by $\pm k_{\alpha x}$ and $\pm k_{\alpha x}^{\prime}$ and correspond to modes $e^{\pm ik_{\alpha x}x}$ and $e^{\pm ik_{\alpha x}^{\prime}x}$, respectively. The subscript $\alpha=L,M,R$ denotes the incident, barrier and transmitting regions, respectively. Next, we discuss the instances of Eq.~(\ref{eq:wave-func}) in different scenarios. As we will show later that modes in the barrier do not affect the way of solving the wavefunction, we classify the scenarios by the modes (in mathematical) in the incident and transmitting regions.

\subsection{Scenario of two propagating modes and two evanescent modes}\label{General situation of one plane wave and one evanescent wave}   

In the scenario that two propagating modes and two evanescent modes (mathematically) exist in both sides of the barrier, the wavefunction Eq.~(\ref{eq:wave-func}) becomes
    \begin{equation}\label{WF (1+1)}  
    \psi\left( x,y \right)  = e^{i k_y y} \times
        \begin{cases}
	   \begin{pmatrix}
        1   \\
        \zeta^{L}_{1}  \\
        \end{pmatrix}  e^{\lambda^{L}_{1} x} + r
	   \begin{pmatrix}
        1   \\
        \zeta^{L}_{2} \\
        \end{pmatrix}  e^{\lambda^{L}_{2} x} + c^\prime
	   \begin{pmatrix}
        1   \\
        \zeta^{L}_{3}  \\
        \end{pmatrix}  e^{\lambda^{L}_{3} x},    &   x<0,  \\
	   a\begin{pmatrix}
        1   \\
        \zeta^{M}_{1}  \\
        \end{pmatrix}  e^{\lambda^{M}_{1} x} + b
	   \begin{pmatrix}
        1   \\
        \zeta^{M}_{2} \\
        \end{pmatrix}  e^{\lambda^{M}_{2} x} + c
	   \begin{pmatrix}
        1   \\
        \zeta^{M}_{3}  \\
        \end{pmatrix}  e^{\lambda^{M}_{3} x} + d
	   \begin{pmatrix}
        1   \\
        \zeta^{M}_{4}  \\
        \end{pmatrix}  e^{\lambda^{M}_{4} x},     &   0 \le x \le D, \\
	   t\begin{pmatrix}
        1   \\
        \zeta^{L}_{1}  \\
        \end{pmatrix}  e^{\lambda^{L}_{1} x}  + d^\prime
	   \begin{pmatrix}
        1   \\
        \zeta^{L}_{4}  \\
        \end{pmatrix}  e^{\lambda^{L}_{4} x},      &   D<x.   \\
        \end{cases}
    \end{equation}
where the superscripts $L, M, R$ denote the incident(left), barrier(medial) and transmitting(right) regions, respectively.
Note since $V(x)$ has the same value for the incident($L$) and transmitting($R$) regions, pseudospin states in the transmitting($R$) region have been  represented by ones in the incident($L$) region. For pseudospin states $\begin{pmatrix}
        1   \\
        \zeta^{L}_{1/2/3/4}  \\
        \end{pmatrix}$, subscript 1(2) corresponds to the propagating wave with positive (negative) component of velocity in the $x$-direction $ v_{x} \equiv \frac{1}{\hbar} \frac{\partial E}{\partial k_x} $, while 3(4) corresponds to the evanescent wave with positive (negative) wavenumber $\lambda_x$ in the $x$-direction, respectively. 

Considering the continuity of both the wavefunction  Eq.~(\ref{WF (1+1)}) and its partial derivative $\partial_x \psi(x,y)  $ respectively at $x=0$, we have $\left[ \psi \left( x,y \right) \right]_{x\to 0^-} = \left[ \psi \left( x,y \right) \right]_{x\to 0^+} $ and $\left[ \partial_x \psi \left( x,y \right) \right]_{x\to 0^-} = \left[ \partial_x \psi \left( x,y \right) \right]_{x\to 0^+} $. Similarly, at $x=D$, we have $ \left[ \psi \left( x,y \right) \right]_{x\to D^-} = \left[ \psi \left( x,y \right) \right]_{x\to D^+} $ and $\left[ \partial_x \psi \left( x,y \right) \right]_{x\to D^-} = \left[ \partial_x \psi \left( x,y \right) \right]_{x\to D^+} $. These four equations can be simplified to be
    \begin{equation}\label{matrix form 1 (1+1)}  
    \begin{aligned}
        P
        \begin{pmatrix}
        1   \\
        r  \\
        c^\prime \\
        \end{pmatrix}  &= 
        M   
	   \begin{pmatrix}
        a   \\
        b   \\
        c   \\
        d  \\
        \end{pmatrix}    \\
    \end{aligned}
    \end{equation}
and
    \begin{equation}\label{matrix form 2 (1+1)}  
    \begin{aligned}
        M  \Lambda
        \begin{pmatrix}
        a   \\
        b  \\
        c  \\
        d  \\
        \end{pmatrix}  &= 
        Q
	   \begin{pmatrix}
        t   \\
        d^\prime   \\
        \end{pmatrix}    \\
    \end{aligned}
    \end{equation}
where $P \equiv \begin{pmatrix}
        1                                       &  1                                       &    1   \\
        \zeta^{L}_{1}                      & \zeta^{L}_{2}                       &  \zeta^{L}_{3} \\
        \lambda^{L}_{1}                    &  \lambda^{L}_{2}                    &  \lambda^{L}_{3}  \\
        \zeta^{L}_{1} \lambda^{L}_{1} & \zeta^{L}_{2} \lambda^{L}_{2}  &  \zeta^{L}_{3} \lambda^{L}_{3}  \\
        \end{pmatrix}  $, $M \equiv \begin{pmatrix}
        1  &  1   &  1   &   1 \\
        \zeta^{M}_{1}         &  \zeta^{M}_{2}       &  \zeta^{M}_{3}      &  \zeta^{M}_{4}  \\
        \lambda^{M}_{1}       &  \lambda^{M}_{2}     &  \lambda^{M}_{3}    &  \lambda^{M}_{4} \\
        \zeta^{M}_{1} \lambda^{M}_{1} & \zeta^{M}_{2} \lambda^{M}_{2}  &  \zeta^{M}_{3} \lambda^{M}_{3}   &  \zeta^{M}_{4} \lambda^{M}_{4}  \\
        \end{pmatrix}  $, 
$\Lambda \equiv \begin{pmatrix}
        e^{\lambda^{M}_{1} D}                     &  0        &    0   &  0 \\
       0                &  e^{\lambda^{M}_{2} D}                       &  0 &  0\\
        0         & 0          &   e^{\lambda^{M}_{3} D}  &  0 \\
        0 & 0  &  0&    e^{\lambda^{M}_{4} D}\\
        \end{pmatrix}  $, $Q \equiv \begin{pmatrix}
        e^{\lambda^{L}_{1} D}        &  e^{\lambda^{L}_{4} D}   \\
        \zeta^{L}_{1} e^{\lambda^{L}_{1} D}  &  \zeta^{L}_{4} e^{\lambda^{L}_{4} D} \\
        \lambda^{L}_{1}  e^{\lambda^{L}_{1} D}    &  \lambda^{L}_{4}     e^{\lambda^{L}_{4} D}      \\
        \zeta^{L}_{1} \lambda^{L}_{1} e^{\lambda^{L}_{1} D}   & \zeta^{L}_{4} \lambda^{L}_{4} e^{\lambda^{L}_{4} D}\\
        \end{pmatrix} $.
The combination of Eq.~(\ref{matrix form 1 (1+1)}) and Eq.~(\ref{matrix form 2 (1+1)}):
    \begin{equation}\label{intermediate relation (1+1)}  
    \begin{aligned}
        M^{-1}    P
        \begin{pmatrix}
        1   \\
        r  \\
        c^\prime \\
        \end{pmatrix}  &=    
	   \begin{pmatrix}
        a   \\
        b   \\
        c   \\
        d  \\
        \end{pmatrix}   =
        \Lambda ^{-1}   M^{-1}
        Q   
	   \begin{pmatrix}
        t   \\
        d^\prime   \\
        \end{pmatrix}    \\
    \end{aligned}
    \end{equation}
is a linear relation from where parameters $r$, $t$ can easily get. Thus, the reflection ($R$) and transmission ($T$) probabilities can be calculated as $R=|r|^2$ and  $T=|t|^2$.

\subsection{Scenario of four propagating modes}\label{General situation of two propagating modes}   

In the scenario that four propagating modes exist in both sides of the barrier, the scattering happens between modes with different magnitudes of velocities should be considered. To change the algorithm as little as possible, we denote $\lambda^{L}_{1} = ik_x$, $\lambda^{L}_{2} = -ik_x$, $\lambda^{L}_{3} = ik_x^\prime$ and $\lambda^{L}_{4} = -ik_x^\prime$. And the wavefunction  Eq.~(\ref{eq:wave-func}) becomes
    \begin{equation}\label{eq:wavefunc4}  
    \psi  \left( x,y \right)  = e^{ik_y y} \times
        \begin{cases} \frac{\displaystyle 1}{\displaystyle \sqrt{|v_{x}| }}  \alpha_{1,2}
	   \begin{pmatrix}
        1   \\
        \zeta^{L}_{1}  \\
        \end{pmatrix}  e^{\lambda^{L}_{1} x} + \frac{\displaystyle r}{\displaystyle \sqrt{|v_{x}|} } \alpha_{1,2}
	   \begin{pmatrix}
        1   \\
        \zeta^{L}_{2} \\
        \end{pmatrix}  e^{\lambda^{L}_{2} x} + \frac{\displaystyle r^\prime} {\displaystyle \sqrt{|v_{x}^\prime|} } \alpha_{3,4}
	   \begin{pmatrix}
        1   \\
        \zeta^{L}_{4}  \\
        \end{pmatrix}  e^{\lambda^{L}_{4} x},    &   x<0,  \\
	   a^\prime \begin{pmatrix}
        1   \\
        \zeta^{M}_{1}  \\
        \end{pmatrix}  e^{\lambda^{M}_{1} x} + b^\prime
	   \begin{pmatrix}
        1   \\
        \zeta^{M}_{2} \\
        \end{pmatrix}  e^{\lambda^{M}_{2} x} + c^\prime
	   \begin{pmatrix}
        1   \\
        \zeta^{M}_{3}  \\
        \end{pmatrix}  e^{\lambda^{M}_{3} x} + d^\prime
	   \begin{pmatrix}
        1   \\
        \zeta^{M}_{4}  \\
        \end{pmatrix}  e^{\lambda^{M}_{4} x},     &   0 \le x \le D, \\
	   \frac{\displaystyle t}{\displaystyle \sqrt{|v_{x}|}}  \alpha_{1,2}
        \begin{pmatrix}
        1   \\
        \zeta^{L}_{1}  \\
        \end{pmatrix}  e^{\lambda^{L}_{1} x}  + \frac{\displaystyle t^\prime}{\displaystyle \sqrt{|v_{x}^\prime|} } \alpha_{3,4}
	   \begin{pmatrix}
        1   \\
        \zeta^{L}_{3}  \\
        \end{pmatrix}  e^{\lambda^{L}_{3} x},      &   D<x.   \\
        \end{cases}
    \end{equation}
where $\alpha_{1,2} = \frac{1}{\sqrt{1+ |\zeta^{L}_{1}|^2}} = \frac{1}{\sqrt{1+|\zeta^{L}_{2}|^2}}$ and  $\alpha_{3,4} = \frac{1}{\sqrt{1+|\zeta^{L}_{3}|^2}} = \frac{1}{\sqrt{1+|\zeta^{L}_{4}|^2}}$ are normalization constants. Comparing with Eq.(\ref{WF (1+1)}), $\begin{pmatrix}
        1   \\
        \zeta^{L}_{3/4}  \\
        \end{pmatrix}$ here are propagating modes and subscript 3(4) corresponds to the positive (negative) component of velocity in the $x$-direction. To simplify the calculation, we rescale the incident wave, resulting Eq.~(\ref{eq:wavefunc4}) to be
    \begin{equation}\label{}  
    \psi \left( x,y \right)  = e^{ik_y y} \times
        \begin{cases}  
	   \begin{pmatrix}
        1   \\
        \zeta_{\text{I},1}  \\
        \end{pmatrix}  e^{\lambda_{\text{I},1} x} +   r
	   \begin{pmatrix}
        1   \\
        \zeta_{\text{I},2} \\
        \end{pmatrix}  e^{\lambda_{\text{I},2} x} + \gamma r^\prime
	   \begin{pmatrix}
        1   \\
        \zeta_{\text{I},4}  \\
        \end{pmatrix}  e^{\lambda_{\text{I},4} x},    &   x<0,  \\
	   a \begin{pmatrix}
        1   \\
        \zeta_{\text{II},1}  \\
        \end{pmatrix}  e^{\lambda_{\text{II},1} x} + b 
	   \begin{pmatrix}
        1   \\
        \zeta_{\text{II},2} \\
        \end{pmatrix}  e^{\lambda_{\text{II},2} x} + c 
	   \begin{pmatrix}
        1   \\
        \zeta_{\text{II},3}  \\
        \end{pmatrix}  e^{\lambda_{\text{II},3} x} + d 
	   \begin{pmatrix}
        1   \\
        \zeta_{\text{II},4}  \\
        \end{pmatrix}  e^{\lambda_{\text{II},4} x},     &   0 \le x \le D, \\
	     t 
        \begin{pmatrix}
        1   \\
        \zeta_{\text{I},1}  \\
        \end{pmatrix}  e^{\lambda_{\text{I},1} x}  + \gamma t^\prime
	   \begin{pmatrix}
        1   \\
        \zeta_{\text{I},3}  \\
        \end{pmatrix}  e^{\lambda_{\text{I},3} x},      &   D<x.   \\
        \end{cases}
    \end{equation}
where $\gamma = \sqrt{\left|\frac{v_{g,x}}{v_{g,x}^\prime}\right|} \frac{\alpha_{3,4}}{\alpha_{1,2}} = \sqrt{\left|\frac{v_{g,x}}{v_{g,x}^\prime}\right| \cdot \frac{1+ |\zeta_{\text{I},1}| ^2}{1+|\zeta_{\text{I},3}|^2}}$. The only difference with Eq.(\ref{WF (1+1)}), in mathematical, is that $c^\prime$ and $d^\prime$ are replaced by $\gamma r^\prime$ and $\gamma t^\prime$, respectively. It indicates that following the same approach of solving Eq.(\ref{WF (1+1)}), parameters $r$, $r'$, $t$, $t'$ can easily get. Thus, the reflection ($R$) and transmission ($T$) probabilities can be calculated as $R=|r|^2+|r'|^2$ and  $T=|t|^2+|t'|^2$.

\subsection{Scenario of normal incidence} \label{Analytical results for normally incidence (the 0-th mode)}

The normal incidence requires $v_{y} \equiv \frac{1}{\hbar} \frac{\partial E}{\partial k_y}$ to be 0. A sufficient condition is $k_y = 0$, which corresponds to the 0-th mode of energy bands. Denote $s=\text{sign}\left(E-V \right)$, then the eigenstate and corresponding wave modes are:
    \begin{equation}\label{}  
        \begin{cases}
	   \displaystyle
        \zeta_{1,2} =  \begin{pmatrix}
           \zeta_A   \\
           \zeta_B   \\
           \end{pmatrix} =     \frac{1}{2}    \begin{pmatrix}
           1-s    \\
           1+s    \\
           \end{pmatrix} , & \lambda_{x1,2} = \pm i\arccos\left(1-\frac{|E-V|}{2} \right), \text{\quad propagating} \\
        \displaystyle 
        \zeta_{3,4} =  \begin{pmatrix}
           \zeta_A   \\
           \zeta_B   \\
           \end{pmatrix} =    \frac{1}{2}        \begin{pmatrix}
           1+s   \\
           1-s   \\
           \end{pmatrix} , & \lambda_{x3,4} = \pm \operatorname{arcosh}\left(1+\frac{|E-V|}{2} \right), \text{\quad evanescent} \\
        \end{cases}
    \end{equation}
The wavefunction  Eq.~(\ref{eq:wave-func}) now is similar to Eq.~(\ref{WF (1+1)}) as:
    \begin{equation}\label{}  
    \psi\left( x,y \right)  =  \frac{e^{ik_y y}}{2}  \times
        \begin{cases}
	   \begin{pmatrix}
        1-s   \\
        1+s   \\
        \end{pmatrix}  e^{\lambda^{L}_{1} x} + r
	   \begin{pmatrix}
        1-s    \\
        1+s \\
        \end{pmatrix}  e^{\lambda^{L}_{2} x} + c^\prime
	   \begin{pmatrix}
        1+s   \\
        1-s   \\
        \end{pmatrix}  e^{\lambda^{L}_{3} x},    &   x<0,  \\
	   a\begin{pmatrix}
        1-s^{\prime}   \\
        1+s^{\prime}  \\
        \end{pmatrix}  e^{\lambda^{M}_{1} x} + b
	   \begin{pmatrix}
        1-s^{\prime}   \\
        1+s^{\prime} \\
        \end{pmatrix}  e^{\lambda^{M}_{2} x} + c
	   \begin{pmatrix}
        1+s^{\prime}  \\
        1-s^{\prime}  \\
        \end{pmatrix}  e^{\lambda^{M}_{3} x} + d
	   \begin{pmatrix}
        1+s^{\prime}   \\
        1-s^{\prime}  \\
        \end{pmatrix}  e^{\lambda^{M}_{4} x},     &   0 \le x \le D, \\
	   t\begin{pmatrix}
        1-s   \\
        1+s  \\
        \end{pmatrix}  e^{\lambda^{L}_{1} x}  + d^\prime
	   \begin{pmatrix}
        1+s   \\
        1-s  \\
        \end{pmatrix}  e^{\lambda^{L}_{4} x},      &   D<x.   \\
        \end{cases}
    \end{equation}
where $s=\text{sign}(E)$ and $s^\prime=\text{sign}(E-V_s )$. Following the same approach, one can get paremeters $r$, $t$ and Eq.~(\ref{eq:transmission-ky0}).

\section{The tunneling conductance varies with the barrier's width}\label{app: tunneling conductance}

 Fig.~\ref{fig:conductance-appendix} shows how the relation between tunneling conductance $G$ and the barrier's height $V_s$ varies as a function of barrier's width $D$. For positive $V_s$, the tunneling conductance has significant oscillation with respect to $V_s$ and the frequency is approximately  in direct proportion to $D$. So do Fig.~\ref{fig:conductance}~(b). This phenomenon indicates that the oscillation arise from the resonances and anti-resonances between opposite propagating waves inside the barrier. For negative $V_s$, the tunneling conductance decreases exponentially no matter the value of $D$, which indicates the reflection is due to the opposite pseudospin orientations between incident and transmitted states.

\begin{figure}
	\centering
	\includegraphics[width=0.35\columnwidth]{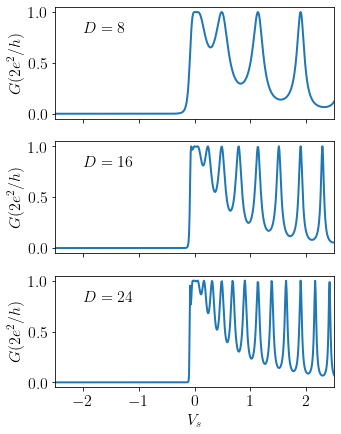} 
	\caption{
		The zero temperature linear tunneling conductance of the checkerboard lattice varies with the barrier's height $V_s$ and width $D$. Incident wave energy $E=-0.1\in[-\Delta,0]$. Other parameters $M=10$, $\Delta=\sqrt{3}(1-\cos\frac{2\pi}{M})\approx0.33$, $D=8$(top panel), $D=16$(middle panel), $D=24$ (bottom panel).}
	\label{fig:conductance-appendix}	
\end{figure}

\end{appendix}

\end{widetext}


\begin{thebibliography}{99}
\bibitem{footnote} According to the anticommutation relation $\{H,\sigma_y\}=0$ and the eigenfunction $H|\bm{k}s\rangle = E_{\bm{k}s} |\bm{k}s\rangle$, one can get $\left\langle\bm{k}s|H\sigma_y+\sigma_y H|\bm{k}s\right\rangle=2E_{\bm{k}s}\langle\bm{k}s|\sigma_y|\bm{k}s\rangle=0$ such that $\langle\bm{k}s|\sigma_y|\bm{k}s\rangle=0$ for $E_{\bm{k}s}\ne0$. Therefore, the pseudospin lies in the ($x$-$z$) plane.
\bibitem{note-fourier} Based on the Taylor expansion, we transform the trigonometric functions in $H_{\tilde{\bm{k}}}$ in the wavevector space into the operators in $\hat{H}_{\bm{k}}$ in the real space with redefining $\bm{k}=\tilde{\bm{k}}-(\pi,\pi)$. For example, for $H_1=\cos(k_x)=1-\frac{k_x^2}{2!}+\frac{k_x^4}{4!}-o(k_x^4)\xrightarrow{\text{Fourier transformation}} \hat{H}_1=1-\frac{1}{2!}(-i\partial_x)^2+\frac{1}{4!}(-i\partial_x)^4-o[(-i\partial_x)^6]$ according to the Fourier transform of derivatives: $(ik)^nf(k)\to\frac{d^n}{dx^n}f(x)$. Then, when the operator $\hat{H}_1$ acts on the wavefunction $\psi(x)=e^{i\lambda x}$, we obtain $\hat{H}_1\psi(x)=[1-\frac{1}{2!}\lambda^2+\frac{1}{4!}\lambda^4-o(\lambda^6)]\psi(x)=\cos(\lambda)\psi(x)$, and here we once again utilize the Taylor expansion.
\end{thebibliography}
\end{document}